\documentclass[aps,prapplied,reprint,superscriptaddress]{revtex4-2}
\usepackage{graphicx}   
\usepackage{mathtools}
\usepackage{mathrsfs}
\usepackage{amsmath}
\usepackage{dcolumn}    
\usepackage{bm}         
\usepackage{hyperref}   
\usepackage[mathlines]{lineno}
\usepackage{float}
\usepackage{xfrac}
\usepackage{soul}
\usepackage[dvipsnames]{xcolor}
\usepackage{nicefrac} 
\begin{document}
\title{Dispersive Readout of a SiMOS Quantum Dot Using a Flip-Chip Integrated Microwave Resonator}
\author{Vo Kim Hieu Van}
\email{h.van@unsw.edu.au}
\affiliation{School of Electrical Engineering and Telecommunications, University of New South Wales, Sydney, NSW 2052, Australia}

\author{Santiago Serrano}
\affiliation{School of Electrical Engineering and Telecommunications, University of New South Wales, Sydney, NSW 2052, Australia}
\affiliation{Diraq, Sydney, NSW 2052, Australia}

\author{C\'edric Boh\'emier}
\affiliation{School of Electrical Engineering and Telecommunications, University of New South Wales, Sydney, NSW 2052, Australia}

\author{Ajit Dash}
\affiliation{School of Electrical Engineering and Telecommunications, University of New South Wales, Sydney, NSW 2052, Australia}
\affiliation{Diraq, Sydney, NSW 2052, Australia}

\author{Fay E. Hudson}
\affiliation{School of Electrical Engineering and Telecommunications, University of New South Wales, Sydney, NSW 2052, Australia}
\affiliation{Diraq, Sydney, NSW 2052, Australia}

\author{Tuomo Tanttu}
\affiliation{School of Electrical Engineering and Telecommunications, University of New South Wales, Sydney, NSW 2052, Australia}
\affiliation{Diraq, Sydney, NSW 2052, Australia}

\author{Chih Hwan Yang}
\affiliation{School of Electrical Engineering and Telecommunications, University of New South Wales, Sydney, NSW 2052, Australia}
\affiliation{Diraq, Sydney, NSW 2052, Australia}

\author{MengKe Feng}
\affiliation{School of Electrical Engineering and Telecommunications, University of New South Wales, Sydney, NSW 2052, Australia}
\affiliation{Diraq, Sydney, NSW 2052, Australia}

\author{Ensar Vahapoglu}
\affiliation{School of Electrical Engineering and Telecommunications, University of New South Wales, Sydney, NSW 2052, Australia}
\affiliation{Diraq, Sydney, NSW 2052, Australia}

\author{Florian K. Unseld}
\affiliation{Diraq, Sydney, NSW 2052, Australia}

\author{Wee Han Lim}
\affiliation{Diraq, Sydney, NSW 2052, Australia}

\author{Andrea Morello}
\affiliation{School of Electrical Engineering and Telecommunications, University of New South Wales, Sydney, NSW 2052, Australia}

\author{Andrew S. Dzurak}
\affiliation{School of Electrical Engineering and Telecommunications, University of New South Wales, Sydney, NSW 2052, Australia}
\affiliation{Diraq, Sydney, NSW 2052, Australia}

\author{Kok Wai Chan}
\email{kokwai@unsw.edu.au}
\affiliation{School of Electrical Engineering and Telecommunications, University of New South Wales, Sydney, NSW 2052, Australia}
\affiliation{Diraq, Sydney, NSW 2052, Australia}

\date{\today}

\begin{abstract}
Heterogeneous integration provides a promising route to combine semiconductor quantum dot devices and superconducting microwave circuits, while allowing each component to be fabricated using an optimized process flow. Here, we demonstrate a flip-chip integrated platform for dispersive readout of silicon metal-oxide-semiconductor (SiMOS) quantum dot devices. A SiMOS double quantum dot chip is bonded to a superconducting aluminum resonator chip using indium bump interconnects to enable microwave coupling to the quantum dot gate. We show that the developed flip-chip process is compatible with cryogenic operation of both the SiMOS device and the superconducting resonator, and demonstrate resonator-based detection of charge transitions in the quantum dot system. The readout signal-to-noise ratio follows a dependence of $\sqrt{t}$ with the integration time, reaching $\mathrm{SNR}=1$ at an integration time of approximately 0.3 ms. These results establish flip-chip bonding as a viable integration approach for SiMOS quantum dot devices operating at both dc and microwave frequencies, with potential applications for resonator-based techniques such as spin-photon coupling.
\end{abstract}

\maketitle

Silicon metal-oxide-semiconductor (SiMOS) quantum dots (QDs) is a promising platform for scalable spin-based quantum information processing owing to their compatibility with established semiconductor manufacturing, long spin coherence times and high-fidelity qubit  operation~\cite{veldhorst2015two, huang2019fidelity, steinacker2025industry, bonen2018cryogenic}. As the number of qubits increases, the requirements for device packaging and footprint for on-chip readout become increasingly demanding. Scalable architectures require compact packaging and readout techniques that can be seamlessly integrated with QD devices~\cite{foxen2018qubit, holman20213d}. Dispersive readout provides a promising solution to this challenge~\cite{petersson2010charge, rossi2017dispersive, granel20263d}. In this approach, a quantum dot gate is coupled to a microwave resonator through flip-chip bonding, to enable charge sensing. The gate electrodes used to electrostatically define the QD can therefore also be used for readout, making the technique attractive for dense silicon spin-qubit arrays.

Flip-chip bonding offers a route for integrating resonator-based readout while keeping the semiconductor-based QD device and superconducting microwave resonator physically separated~\cite{holman20213d}. By fabricating the SiMOS QD device and the superconducting resonator on separate chips, each component can be independently optimized before assembly. Here, we demonstrate a flip-chip integrated SiMOS double quantum dot chip bonded to a superconducting aluminium resonator using indium bump interconnects. We subsequently show resonator-based detection of the QD charge transitions and investigate the robustness of the heterogeneously integrated device as a function of temperature and magnetic field.

\begin{figure}[ht!]
\includegraphics[width=0.5\textwidth]{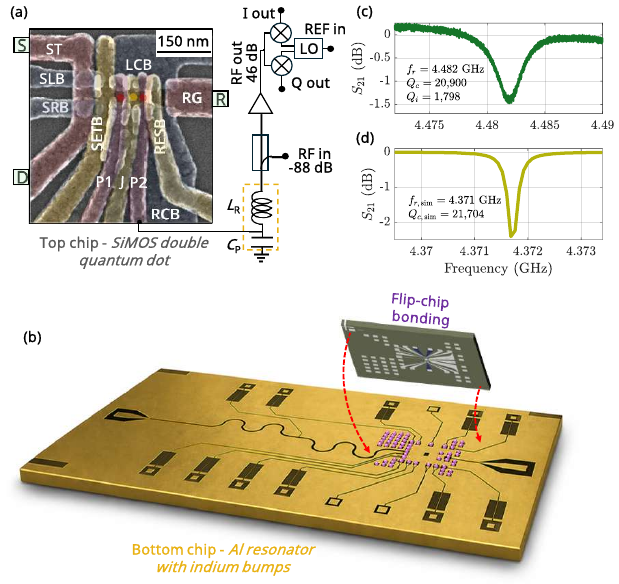}
\caption{\label{fig:device_architecture}
Flip-chip integrated SiMOS quantum dot-resonator architecture. (a) False-coloured scanning electron micrograph of the three-layer aluminum gate-stack architecture SiMOS double quantum dot device. The first, second and third aluminum gate layers
are shown in blue, red and yellow, respectively. (b) The quantum dot chip is flip-chip bonded to a  superconducting aluminium resonator chip using indium bump interconnects. The readout circuit consists of an on-chip superconducting resonator with effective capacitance ($C_P$) and inductance ($L_R$) coupled to the quantum dot gate (P2), with the reflected microwave signal amplified and demodulated using an IQ mixer. (c) The measured and (d) simulated microwave transmission.}
\end{figure}

\begin{figure*}[ht!]
\includegraphics[width=\textwidth]{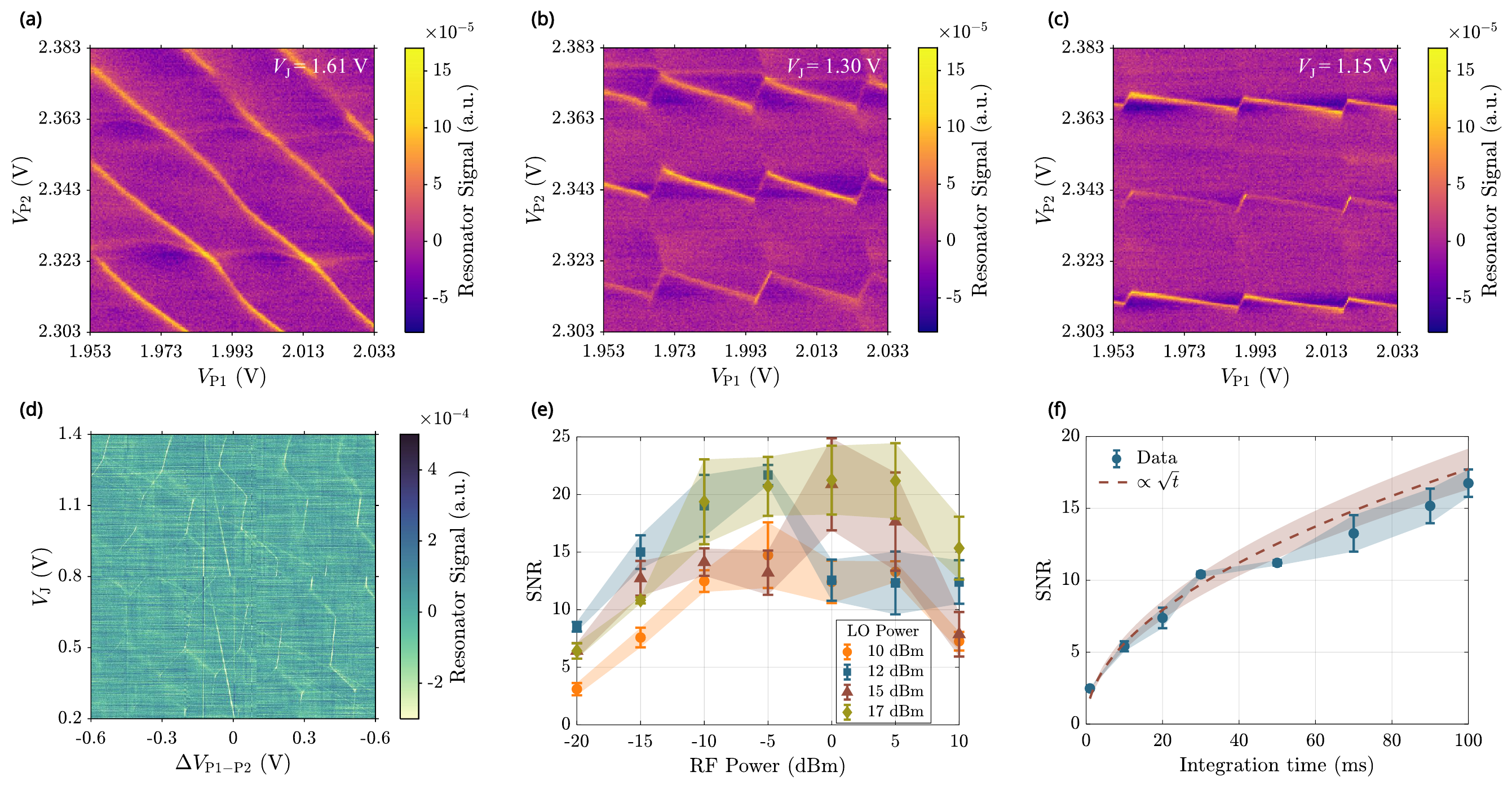}
\caption{\label{fig:dispersive_readout}
Dispersive readout of the flip-chip integrated SiMOS double quantum dot. (a-c) Resonator response in reservoir mode measured as a function of the plunger gate voltages $V_{\mathrm{P1}}$ and $V_{\mathrm{P2}}$, showing charge transitions of the SiMOS quantum dot system detected through the superconducting microwave resonator. (d) Resonator response in isolated mode as a function of the exchange gate voltage ($V_\mathrm{J}$) and the detuning voltage $(\Delta V_{\mathrm{P1-P2}})$, showing repeated charge transitions over an extended voltage range. 
(e) Signal-to-noise ratio (SNR) as a function of applied microwave power for different local oscillator (LO) powers. (f) SNR as a function of integration time. The dashed line shows a $\sqrt{t}$ dependence.}
\end{figure*}

The QD device was fabricated on an intrinsic silicon substrate with a $8~\mathrm{nm}$-thick thermally grown SiO$_2$ gate oxide. The electrostatic gates were defined using a three-layer aluminum gate-stack architecture~\cite{chan2018assessment}, as shown in Fig.~\ref{fig:device_architecture}(a). The QDs are electrostatically defined beneath the two plunger gates, P1 and P2, with the exchange gate, J, controlling the interdot tunnel coupling. Lateral confinement is provided by the LCB and RCB gates, while SETB, RESB, and RG are used to tune the tunnel coupling between the QDs and the S, D and R electron reservoirs. Although the device also includes an on-chip single-electron transistor charge sensor, this sensor was not used in the measurements reported here.

The superconducting microwave circuit was fabricated on a separate chip. The aluminum resonator, shown in 3D render in Fig.~\ref{fig:device_architecture}(b), was deposited by electron-beam evaporation. The resonator was designed such that, after bonding, it will be capacitively connected to the P2 plunger gate of the SiMOS QD device, while the remaining gates are routed to fan-out pads on the resonator chip. A Ti/Pd ($5~\mathrm{nm}$/$20~\mathrm{nm}$) layer was deposited as under bump metallization (UBM) on the resonator chip. These indium bumps, shown as the purple structures in Fig.~\ref{fig:device_architecture}(b) were subsequently deposited on these UBM on the resonator chip. A total of 19 indium bumps were used to provide electrical connections between the resonator and the SiMOS QDs chips. In addition, 40 electrically floating bumps were added to improve the mechanical stability of the bonded stack and to support the top chip during bonding and thermal cycling.

The SiMOS chip formed the top chip, with a lateral size of $2 \times 2~\mathrm{mm}^2$, while the resonator chip formed the bottom chip, with a lateral size of $10 \times 6~\mathrm{mm}^2$. The two chips were assembled by thermocompression flip-chip bonding process used in Ref.~\cite{van2026development}.

The packaged chip was cooled to 60 mK for electrical and microwave characterization. A simplified circuit schematic of the reflectometry setup is shown in Fig.~\ref{fig:device_architecture}(a). The superconducting resonator is represented by an effective inductance $L_R$ and capacitance $C_P$, which define the resonator response coupled to the P2 plunger gate of the SiMOS DQD. The microwave excitation was supplied by a microwave source and applied to the resonator input after a total attenuation of $-88~\mathrm{dB}$ along the input line. The output signal from the resonator was amplified by a total gain of $46~\mathrm{dB}$ before being demodulated using an IQ mixer. A reference signal from the microwave source was used as the local oscillator (LO) input to the mixer. The resulting in-phase ($I$) and quadrature ($Q$) components were recorded using an FPGA-based acquisition system.

The resonator response was measured as a function of microwave frequency after flip-chip bonding. As shown in Fig.~\ref{fig:device_architecture}(c), the resonator exhibits a resonance at $f_r = 4.482~\mathrm{GHz}$. From a fit to the complex resonator response using the real and imaginary components of the measured signal~\cite{Probst2015FitRes}, we extract an internal quality factor $Q_i = 1798$ and a coupling quality factor $Q_c = 20900$. These values confirm that the superconducting resonator remains operational after flip-chip bonding and that microwave coupling to the SiMOS device is preserved. To understand the origin of the measured resonator response, a CST Studio Suite High Frequency simulation was done. The simulation method is limited insofar as $Q_i$ is governed by processes that are not captured in this methodology, such as two-level fluctuators in the oxide \cite{Gao2008TLS}. However, we can retrieve an estimate of $Q_c$ and $f_r$. We report $f_{r, \, sim}=4.371$ GHz and $Q_{c, \, sim}=21704$, both of which are on the same order of magnitude as the measured data. The simulation also provides an electric field distribution and confirms that the compound mode does have an anti-node at the quantum dots.

We measured the resonator response of the integrated device while sweeping the plunger gate voltages $V_{\mathrm{P1}}$ and $V_{\mathrm{P2}}$. Figures~\ref{fig:dispersive_readout}(a)--\ref{fig:dispersive_readout}(c) show the charge stability maps of the SiMOS DQD measured in the reservoir-coupled mode~\cite{lai2011pauli}, with the device operated in the many-electron regime. To remove slowly varying background offsets, the data were subtracted with the mean value of each horizontal trace. In this configuration, the QDs electrostatically defined under P1 and P2 remain coupled to the electron reservoirs, allowing charge transitions associated with electron loading and unloading to be detected as changes in the resonator response. The measurements also demonstrate control of the interdot tunnel coupling using the exchange gate, J. In Fig.~\ref{fig:dispersive_readout}(a), the DQD is operated in a strongly interdot-coupled regime. As the interdot tunnel coupling is reduced, the charge stability pattern evolves into the intermediate-coupling regime shown in Fig.~\ref{fig:dispersive_readout}(b). In the weakly coupled regime shown in Fig.~\ref{fig:dispersive_readout}(c), the charge transitions evolve into a checkerboard pattern, due to reduced coupling between the DQD. These measurements show that the device can be tuned across different interdot coupling regimes while maintaining clear dispersive readout sensitivity.

Figure~\ref{fig:dispersive_readout}(d) shows the resonator response in a regime where the DQD is largely decoupled from the electron reservoirs~\cite{yang2020operation, han20252}. In this isolated regime, the total electron number is fixed over the measured voltage range, and the observed transitions correspond to charge rearrangements within the isolated system. Multiple charge-transition lines are visible as a function of the detuning voltage $\Delta V_{\mathrm{P1-P2}}$ and the exchange gate voltage $V_{\mathrm{J}}$, confirming that the flip-chip integrated resonator can detect charge motion even in the isolated mode operation. These measurements demonstrate that the flip-chip process preserves both of the dc gate control and microwave charge sensitivity in the SiMOS DQD device.

To quantify the charge-readout performance, we extracted the signal-to-noise ratio (SNR) from a single interdot charge transition between the P1 and P2 QDs ~\cite{zheng2019rapid, west2019gate}. The SNR was determined from one-dimensional cuts through the resonator response magnitude, where the in-phase and quadrature components were combined as $\sqrt{I^2+Q^2}$. For each trace, a baseline region away from the transition was first selected to estimate the background signal and noise level. The charge-transition feature was then fitted using a Gaussian peak on top of the local baseline. The signal amplitude was defined as the fitted peak height relative to the baseline, while the noise amplitude was taken as the standard deviation of the baseline fluctuations to determine the SNR. In Fig.~\ref{fig:dispersive_readout}(e), this procedure was repeated for different applied microwave input powers and LO powers. All measurements were performed with an integration time of $50~\mathrm{ms}$. Each data point represents the mean SNR extracted from three repeated measurements taken under the same conditions on the same interdot charge transition, and the error bars represent the corresponding variation between these measurements.

\begin{figure*}[!t]
\includegraphics[width=\textwidth]{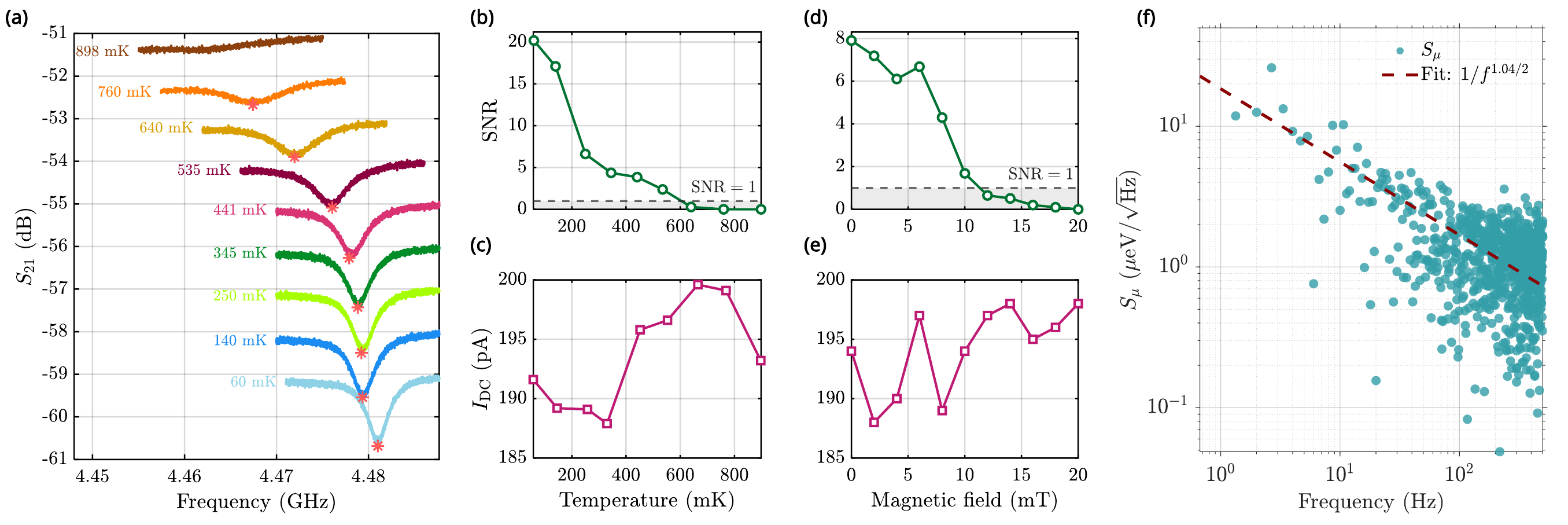}
\caption{\label{fig:temperature_field_noise}
Cryogenic performance and noise characterization of the flip-chip integrated dispersive readout device. (a) Microwave transmission $S_{21}$ of the superconducting resonator measured at different temperatures from $60~\mathrm{mK}$ to $898~\mathrm{mK}$. The resonator response remains visible up to 760 mK, with the resonance dips marked by asterisks. The data are vertically shifted for clarity. (b, d) Charge-readout SNR as a function of temperature and in-plane magnetic field, respectively. The dashed lines indicate $\mathrm{SNR}=1$. (c, e) The corresponding dc transport current through the SiMOS device as a function of temperature and magnetic field, respectively, showing that the flip-chip bonded device remains electrically functional over the same range. (f) Electrochemical-potential noise amplitude spectral density $S_{\mu}$ extracted from the dispersive readout signal. The dashed line shows a $1/f^{1.04/2}$ fit, indicating approximately $1/f$ charge-noise behaviour. 
The charge-noise amplitude at $1~\mathrm{Hz}$ is $134.01~\mu\mathrm{eV}/\sqrt{\mathrm{Hz}}$.}
\end{figure*}

At low microwave input power, the SNR is limited by the small resonator response generated by the interdot charge transition. Increasing the microwave input power initially improves the SNR by increasing the dispersive signal. However, at higher microwave input powers the SNR no longer improves and eventually decreases. This behavior is consistent with power broadening of the charge transition, where strong microwave excitation broadens the transition~\cite{west2019gate}. The dependence on LO power reflects the conversion efficiency of the IQ demodulation. In the present setup, LO power of $12~\mathrm{dBm}, 15~\mathrm{dBm}$, and $17~\mathrm{dBm}$ provided high SNR and a robust operating window for resonator drive powers between $-10~\mathrm{dBm}$ and $5~\mathrm{dBm}$, with SNR $~\mathrm{>}~10$.

We measured the dependence of SNR on integration time using an interdot charge transition~\cite{zheng2019rapid, west2019gate} and the same SNR extraction method. In this experiment, an LO power of $12~\mathrm{dBm}$ and a resonator drive power of $-5~\mathrm{dBm}$ were selected based on the previously identified robust operating regime. For each integration time, three repeated measurements were used to calculate the mean SNR and the corresponding error bars. As shown in Fig.~\ref{fig:dispersive_readout}(f), the SNR increases with integration time and follows an approximately $\sqrt{t}$ dependence. This scaling is expected when the measurement is limited predominantly by white noise, since averaging for a longer integration time reduces the noise amplitude. A fit to the form of $\mathrm{SNR} \propto \sqrt{t}$ gives an estimated integration time of approximately 0.3~ms at $\mathrm{SNR}=1$.

Next, we investigated the robustness of the flip-chip integrated dispersive readout device as a function of temperature and magnetic field. Since the microwave resonator is made of aluminium, its superconducting properties are expected to be sensitive to both temperature and magnetic field, with a critical temperature of order $1~\mathrm{K}$ and a critical magnetic field of order $10~\mathrm{mT}$~\cite{khukhareva1963superconducting,Sandor1965critical}. We monitor the resonator response while increasing the device temperature from $60~\mathrm{mK}$ to $898~\mathrm{mK}$, as shown in Fig.~\ref{fig:temperature_field_noise}(a). As the temperature approaches the critical-temperature range of aluminium, the resonator response weakens. A clear resonance is observed up to $760~\mathrm{mK}$, beyond which the resonance could not be resolved. We observed that the resonance frequency decreases with increasing temperature.

To quantify how the change in resonator response affects the charge sensing, we extracted the charge-readout SNR from the two-dimensional charge-stability maps measured at different temperature. Unlike the one-dimensional peak-fitting procedure used in Fig.~\ref{fig:dispersive_readout}(e,f), here the SNR is obtained directly from the two-dimensional resonator-response data. First, the in-phase and quadrature components are combined as $\sqrt{I^2+Q^2}$ after background subtraction. Then, the signal from a region of interest is selected on a visible charge-transition line, while a reference region away from the charge transition is used to estimate the background noise. The signal and noise distributions are fitted with Gaussian functions, and the SNR is extracted from the separation between the signal and reference distributions relative to their widths.

Figure~\ref{fig:temperature_field_noise}(b) shows that the charge-readout SNR decreases with increasing temperature. The SNR is highest at base temperature and falls below 1 at $T > 640~\mathrm{mK}$. This loss of charge sensitivity is attributed primarily to the degradation of the superconducting resonator response and the increased thermal broadening of the charge transition. In contrast, the dc transport current between the source and reservoir ohmics remains visible within fluctuations over the same temperature range, as shown in Fig.~\ref{fig:temperature_field_noise}(c). We note that the indium bumps used in the flip-chip process has a superconducting critical temperature of approximately 3.4 K~\cite{foxen2018qubit, ho1972thermal} and should not affect their function as electrical interconnects.

A similar characterization was performed as a function of magnetic field. Figure~\ref{fig:temperature_field_noise}(d) shows the extracted charge-readout SNR as the in-plane $[110]$ magnetic field is increased up to $20~\mathrm{mT}$. The applied field is parallel to the plane of the resonator and orthogonal to the length of the indium bumps. The SNR remains well above 1 at a low magnetic field of up to $10~\mathrm{mT}$. This behavior is consistent with the critical magnetic field of the aluminium resonator. The corresponding dc transport current, shown in Fig.~\ref{fig:temperature_field_noise}(e), remains measurable across the same field range. These measurements show that the functionality of the flip-chip bonding is preserved under conditions where the microwave resonator response becomes the limiting factor for dispersive readout.

Finally, we use the dispersive readout signal to estimate the electrochemical-potential noise spectrum of the integrated system~\cite{wilson2025fast}. A time trace of the resonator response is recorded along a diagonal sweep in P1-P2 gate space near an interdot charge transition. The slope of the averaged resonator response signal, $dI_\mathrm{out}/d(V_{\mathrm{P1}}-V_{\mathrm{P2}})$, where $I_\mathrm{out}$ is the in-phase component, used to convert fluctuations in the measured resonator response into an equivalent noise. The spectrum is obtained from the Fourier transform of the time-domain trace and fitted to the power-law, $S_{\mu}(f)=A/f^{\beta/2}$. As shown in Fig.~\ref{fig:temperature_field_noise}(f), the measured spectrum follows an approximately $1/f$ type dependence, with $\beta \approx 1.04$. The extracted electrochemical-potential noise amplitude at $1~\mathrm{Hz}$ is $18.43~\mu\mathrm{eV}/\sqrt{\mathrm{Hz}}$.

In summary, we have demonstrated that flip-chip bonding can be used to heterogeneously integrate a SiMOS QD device with a separately fabricated superconducting resonator for dispersive readout. We observed that the charge-readout SNR decreases with increasing temperature and magnetic field, consistent with the superconducting properties of the aluminium resonator, while transport through the SiMOS device remains measurable over the same range. These results show that the flip-chip bonded SiMOS-resonator assembly remains operable under these conditions, establishing a viable integration route for large-scale SiMOS quantum dot devices that combine dc transport with microwave-frequency readout. Future applications of this technique include resonator-based approaches such as fast charge sensing and spin-photon coupling.

\section*{Author Contributions}
V.K.H.V fabricated the superconducting resonator and performed flip-chip bonding under the supervision of K.W.C. K.W.C and F.H fabricated the quantum device under the supervision of A.S.D. V.K.H.V performed all measurements and calculations under the supervision of K.W.C, with inputs from S.S, A.D, C.B, T.T, C.H.Y, and E.V. C.B assisted with CST Studio Suite simulations. S.S assisted with the reflectometry and experimental setup. A.D assisted with the experimental setup. V.K.H.V, S.S, C.B, A.D, F.H, T.T, C.H.Y, M.F, E.V, F.K.U, W.H.L, A.M, A.S.D and K.W.C contributed to data interpretation. V.K.H.V wrote the manuscript with inputs from all authors.

\section*{Acknowledgments}
We thank Andrii Torgovkin for assistance with the cryogenic setup. We thank Hemendra Kala and Mariusz Martyniuk for assistance with indium deposition. We acknowledge support from the Australian Research Council (Grants No. FL190100167, CE170100012, and IM230100396), and the U.S. Army Research Office (W911NF-17-1-0198 and W911NF-23-10092). We thank the Australian National Fabrication Facility at the University of New South Wales, University of Western Australia, and the Sydney Nano Foundry, University of Sydney. The authors acknowledge the Defence Trailblazer Quantum Packaging Facility at the Commonwealth Scientific and Industrial Research Organisation in Lindfield, NSW.

\bibliography{References.bib}
\end{document}